\documentclass[a4paper,11pt]{article}    
\pdfoutput=1 
\usepackage{jheppub}
\usepackage{epsfig}
\usepackage[T1]{fontenc} 

\def\arXiv#1{\href{http://arxiv.org/abs/#1}{arXiv:#1}}
\def\arXiv#1#2{\href{http://arxiv.org/abs/#1}{arXiv:#1}}

\def\doi#1#2{\href{http://doi.org/#1}{#2}}

\def\be{\begin{equation}}
\def\ee{\end{equation}}
\def\bea{\begin{eqnarray}}
\def\eea{\end{eqnarray}}
\newcommand{\nn}{\nonumber}

\newcommand\para{\paragraph{}}

\begin{document}

\unitlength = .8mm

\vspace{.5cm}

\title{Topological nodal line semimetals in holography}

\author[a]{Yan Liu}
\author[b,c,d]{, Ya-Wen Sun}
\affiliation[a]{Department of Space Science, and International Research Institute
of Multidisciplinary Science, 
\\Beihang University,  Beijing 100191, China} 
\affiliation[b]{School of physics  \& CAS Center for Excellence in Topological Quantum Computation, University of Chinese Academy of Sciences, Beijing 100049, China}
\affiliation[c]{Kavli Insititute for Theoretical Sciences, University of Chinese Academy of Sciences, Beijing 100049,  China}
\affiliation[d]{CAS Key Laboratory of Theoretical Physics, Institute of Theoretical Physics, 
\\Chinese Academy of Sciences, Beijing 100190, China}
\emailAdd{yanliu@buaa.edu.cn, yawen.sun@ucas.ac.cn}

\vspace{.2cm}

\abstract{
We show a holographic model of a strongly coupled topological nodal line semimetal (NLSM) and find that the NLSM phase could go through a quantum phase transition to a topologically trivial state. The dual fermion spectral function shows that there are multiple Fermi surfaces each of which is a closed nodal loop in the NLSM phase. The topological structure in the bulk is induced by the IR interplay between the dual mass operator and the operator that deforms the topology of the Fermi surface. We propose a practical framework for building various strongly coupled topological semimetals in holography, which indicates that at strong coupling topologically nontrivial semimetal states generally exist.}


\maketitle

\section{Introduction}

During the last decade, our understanding towards topological states of matter has grown enormously in condensed matter physics. 
Examples of topological states of matter include topological insulators, topological Weyl  semimetals (WSMs)/NLSMs, anomalous Hall states, and topological superconductors (see e.g. \cite{Witten:2015aoa, review-tsm} for a review). 
Most properties of the topological systems that we have known were built on the weakly coupled picture. An important challenge in the understanding  
of topological states of matter is to consider the effect of interactions, especially strong interactions, on the topological structures of these systems. Weakly coupled nontrivial topological structure could be destroyed by strong interactions or new strongly coupled topological states of matter could arise. This has been a rapidly growing research area in condensed matter during the recent years. However, it remains quite difficult to attack this problem due to the lack of band structure and no notion of quasiparticles and technical difficulty at strong coupling in the condensed matter physics, especially for gapless topological systems. 
Some attempts in this direction could be found in \cite{interaction1, interaction2}.

 \para
To study strongly coupled topological states of matter, a very powerful tool from string theory, the AdS/CFT correspondence, which maps a  $d+1$ dimensional strongly coupled field theory to a $d+2$ dimensional weakly coupled classical gravitational theory, would be extremely helpful.  As a strong-weak duality, AdS/CFT has obtained lots of success in its applications to strongly coupled condensed matter systems \cite{book, book0, review}. It is, however, still a relatively undeveloped research area to incorporate topological states of matter into the holographic dictionary. Some previous attempts could be found in e.g. \cite{HoyosBadajoz:2010ac, Kristjansen:2016rsc, Seo:2017oyh} for topological insulator and \cite{WeiLi, Bergman:2010gm} for quantum Hall states. Here we shall explore what holography can tell about the strongly interacting topological semimetals. 
 \para
An important question is if there exists a general framework for holographic topological states of matter and to find the corresponding bulk  topological structure. As a first step to answer these questions and inspired by our previous work  \cite{Landsteiner:2015pdh} on a holographic model of strongly coupled WSM state\footnote{
The holographic model also gave an important prediction for the transport property of the system which could be used to detect mixed axial-gravitational anomaly in laboratories \cite{Landsteiner:2016stv}. More study on this holographic model could be found in
\cite{Landsteiner:2015lsa, Copetti:2016ewq, Grignani:2016wyz, Ammon:2016mwa}.}, we will start from building a holographic model of topological NLSM, which is a gapless topological state of matter whose Fermi surface forms closed loops in momentum space. We will show its dual fermion spectral function behavior and reveal the common mathematical structure of topology in the gravitational solutions that the NLSMs and WSMs share. In particular, this bulk topological structure could be generalized to a general paradigm in holography to describe strongly coupled gapless topological states of matter.

\para
In the following of this paper we will first give an example of a new entry in the
holographic dictionary of topological states of matter: the holographic topological nodal line semimetal in Sec. \ref{sec2}. Then the behavior of the corresponding fermion spectral functions of the nodal line semimetal will be shown in Sec. \ref{sec3}. In Sec. \ref{sec4} 
we point out the general bulk topological structure of topological semimetal states. Appendices \ref{secaa}, \ref{secfe}, \ref{secab} contain the equations of motion, free energy calculations of the system and the discussion at the probe limit.

\section{Holographic topological nodal line semimetals} 
\label{sec2} 

A NLSM \cite{burkov1} has a nontrivial shape of Fermi surface where Fermi nodal points form a circle under certain symmetries, e.g. mirror reflection symmetry (see \cite{rev1} for a review). A topologically nontrivial NLSM cannot be gapped by small perturbations unless passing through a topological phase transition to a trivial state. 

 \subsection{A field theoretical model} 

To get some hint for building the dictionary of this model, we first take a look at a simple weakly coupled field theory model that represents a nodal line semimetal. 
The Lagrangian is  
\be\label{eq:1}
\mathcal{L}=i\bar{\psi}\big(\gamma^\mu\partial_\mu-m-\gamma^{\mu\nu} b_{\mu\nu}\big)\psi
\ee
where $\gamma^{\mu\nu}=\frac{i}{2}[\gamma^\mu, \gamma^\nu]\,$ and $b_{\mu\nu}=-b_{\nu\mu}$ is an antisymmetric two form field. Note that to be consistent with the gravity calculations, we take the $(-,+,+,+)$ signature. 
We turn on a nonzero constant $b_{xy}$ component of the two form field and the energy spectrum of the eigenstates of this system are therefore
\be
E_\pm=\pm \sqrt{k_z^2+\Big(2b_{xy}\pm\sqrt{m^2+k_x^2+k_y^2}\,\Big)^2}\,.
\ee 
For $m^2< 4b_{xy}^2$, the system is a topological nodal line semimetal with Fermi points forming a connected circle with radius $\sqrt{4 b_{xy}^2-m^2}$ in the momentum space. In this parameter regime we can see that the system cannot be gapped by small perturbations like an ordinary Dirac field, i.e. a small perturbation in $m$ cannot gap this system. For $m^2 > 4b_{xy}^2$,  the system becomes an insulator and $m^2= 4b_{xy}^2$ is the quantum transition point of this topological phase transition. In the nodal line phase, we see that close to the Fermi line, the dispersion behaves as linear in $\sqrt{k_x^2+k_y^2}-\sqrt{4 b_{xy}^2-m^2}$ with velocity $\sqrt{1-\frac{m^2}{4b_{xy}^2}}$ when $k_z=0$ and linear in $k_z$ with velocity $1$ when $\sqrt{k_x^2+k_y^2}=\sqrt{4 b_{xy}^2-m^2}$.

\begin{figure}[h!]
\begin{center}
\includegraphics[width=0.4\textwidth]{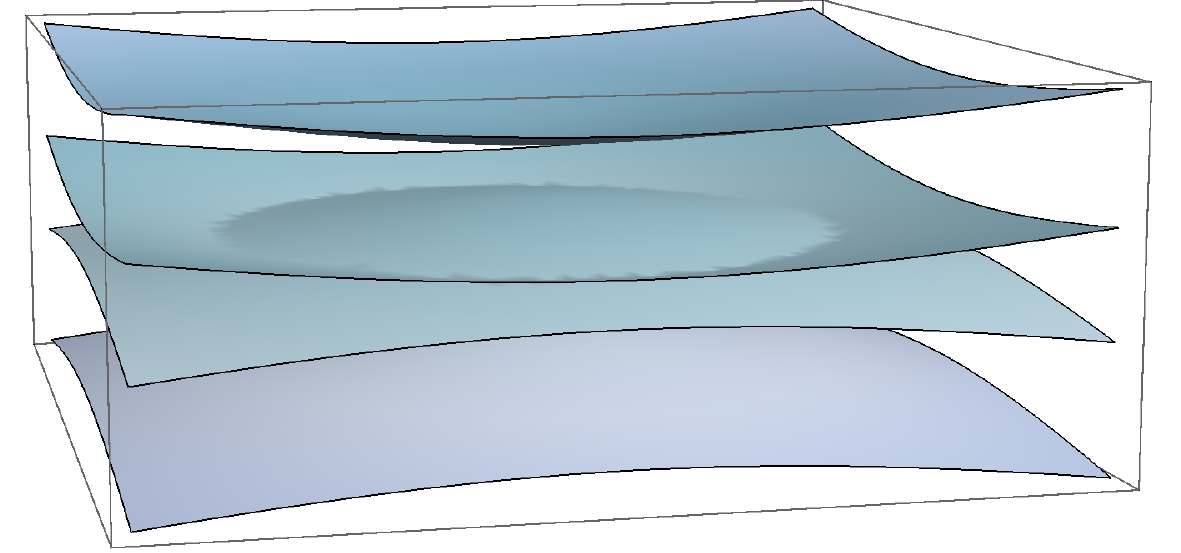}
\includegraphics[width=0.4\textwidth]{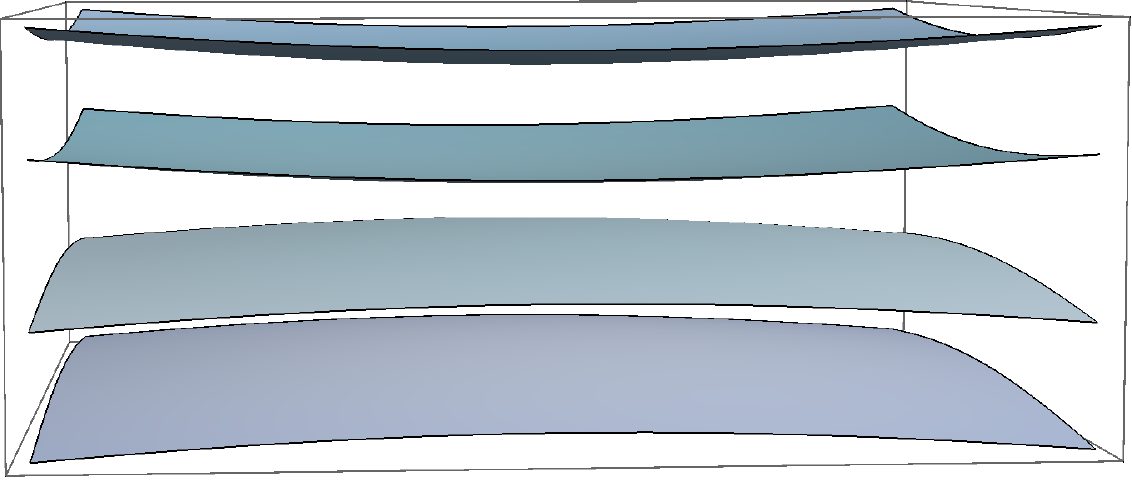}
\end{center}
\vspace{-0.3cm}
\caption{\small The energy spectrum as a function of $k_x, k_y$ for $k_z=0$. Left: there is a nodal line at the band crossing when $m^2< 4b_{xy}^2$. Right: for $m^2 > 4b_{xy}^2$ the system is gapped.}
\label{fig:phase}
\end{figure}
 \para
We can see that the phase described by (\ref{eq:1}) for $m^2< 4b_{xy}^2$ should be a topologically nontrivial nodal line semimetal. Other components of $b_{\mu\nu}$ would play a similar role and produce a nodal line semimetal with slight difference in the spectrum, e.g. $b_{tz}$ would produce an accidental nodal line semimetal.
 \para
The new two form field $b_{\mu\nu}$ term does not break the conservation of the eletric current $J^\mu=\bar{\psi} \gamma^\mu \psi$ while does not conserve the axial current $J_5^\mu=\bar{\psi} \gamma^\mu \gamma^5\psi$. We have the following conservation equations
\bea\label{conservation}
\partial_\mu J^\mu&=&0\,,\\
\label{conservation2}
\partial_\mu J_5^\mu&=&-2m\bar{\psi}\gamma^5\psi-2b_{\mu\nu}\bar{\psi}\gamma^{\mu\nu}\gamma^5\psi\,,
\eea where we have ignored the anomaly terms.
 \para
Meanwhile, let us elaborate more on the differences between Weyl Semimetal (WSM) and nodal line semimetal (NLSM) which are two different novel examples of gapless topological states of matter. In the former case two Weyl points are separated at a distance $b$ while the latter Weyl points form a 1D circle. The low energy effective field theory for weakly coupled WSM and NLSM are different, although both of them can be described as a Dirac fermion coupled to an external field. In the former case, a time reversal symmetry breaking parameter axial $A_z$ is crucial, while for the latter, a two form effective field $b_{xy}$ is crucial which breaks both time reversal and charge conjugate.

\subsection{The holographic model}
A weakly coupled NLSM model is shown in the above subsection, where a coupling term $\bar{\psi}\gamma^{\mu\nu}\psi$ is responsible for deforming the Fermi point to a closed loop. It is a new nontrivial topological semimetal phase compared to WSM in which the Fermi point is seperated to two Weyl points. 
Motivated by the weakly coupled mechanism, we utilize a massive two form field $B_{ab}$ to be dual to an antisymmetric operator which is though not exactly the $\bar{\psi}\gamma^{\mu\nu}\psi$ operator and introduce an axially charged scalar field $\Phi$ to be dual to the operator $\bar{\psi}\psi$ whose source term breaks the axial symmetry and represents the gap effect. We consider\footnote{We set $2\kappa^2=L=1.$ We use $a, b=t, x, y, z, r$ to denote bulk indices, $\mu, \nu=t, x, y, z$ for boundary indices.}
\bea
S&=&\int d^5x\sqrt{-g}\bigg[\frac{1}{2\kappa^2}\bigg(R+\frac{12}{L^2}\bigg)-\frac{1}{4}\mathcal{F}^2-\frac{1}{4}F^2+\frac{\alpha}{3}\epsilon^{abcde}A_a \bigg(3\mathcal{F}_{bc}\mathcal{F}_{de}+F_{bc}F_{de}\bigg)
\nn\\
&&-(D_a \Phi)^*(D^a\Phi)-V_1(\Phi)-\frac{1}{3\eta}\big(\mathcal{D}_{[a}B_{bc]}\big)^*\big(\mathcal{D}^{[a}B^{bc]}\big)
-V_2(B_{ab})-\lambda|\Phi|^2B_{ab}^*B^{ab}\bigg]
\eea
where $\mathcal{F}_{ab}=\partial_a V_b-\partial_b V_a$ is the vector gauge field strength\footnote{Note that a semimetal state is a zero density state so that the vector gauge field will be turned off throughout the calculations in this paper. Here we keep the vector gauge field terms in the action for future convenience in the calculation of transport coefficients.}, $F_{ab}=\partial_a A_b-\partial_b A_a$ is the axial gauge field strength, $D_a=\nabla_a -iq_1 A_a$, $\mathcal{D}_a=\nabla_a -iq_2 A_a$ and
\bea
\mathcal{D}_{[a}B_{bc]}&=&\partial_a B_{bc}+\partial_b B_{ca}+\partial_c B_{ab}
\nonumber\\
&&~
-iq_2 A_a B_{bc}-iq_2 A_b B_{ca}-iq_2 A_c B_{ab}\,.
\eea
$B_{ab}$ is also axially charged as 
the dual operator explicitly breaks the axial symmetry. The potential terms are
\bea
V_1=m_1^2 |\Phi|^2+\frac{\lambda_1}{2} |\Phi|^4\,,~~~~
V_2=m_2^2 B^*_{ab}B^{ab}\,,
\eea 
where $m_1$ and $m_2$ are the mass parameters of the scalar and the two form field. $B_{ab}$ has a mass term because it does not correspond to a conserved operator and we will turn on the $B_{xy}$ component in the following. The $\lambda$ term denotes the interaction effect between the operators $\bar{\psi}\psi$ and the antisymmetric operator, which 
is important to the existence of the topological structure and the topological phase transition. 
Similar to the holographic WSM \cite{Landsteiner:2015pdh}, we introduce the $\lambda_1$ term because it is not possible to 
find nonsingular solutions without this term. $\lambda_1$ characterises 
the number of UV degrees of freedom that does not gap in the IR.  
This indicates that this simple holographic system cannot be completely gapped in the IR and could only be partially gapped at most. The equations of motion, choice of parameters and current conservation equations can be found in appendix \ref{secaa}. 
\para
From the weakly coupled theory, we see that the operators $\bar{\psi}\psi$ and $\bar{\psi}\gamma^5\psi$ correspond to the real and imaginary parts of the complex scalar field in the bulk and the operators $\bar{\psi}\gamma^{\mu\nu}\psi$ and $\bar{\psi}\gamma^{\mu\nu}\gamma^5\psi$ should have similar properties. However, the operators $\bar{\psi}\gamma^{\mu\nu}\psi$ and $\bar{\psi}\gamma^{\mu\nu}\gamma^5\psi$ not independent indicating that the dual field $B_{ab}$ should have a self-dual property. To avoid this problem, here we instead consider an operator different from $\bar{\psi}\gamma^{\mu\nu}\psi$ which nevertheless is antisymmetric in the two indices and has no self duality properties. Note that a previous holographic model considered the self-duality effect \cite{Alvares:2011wb}.\footnote{
Other interesting application of (massless) two form field in holography can be found in \cite{Grozdanov:2017kyl,{Hofman:2017vwr}} for holographic descriptions of 3+1D magnetohydrodynamics and in \cite{Cai:2014oca} for magnetization transitions in AdS$_4$/CFT$_3$.}
 \para
\noindent {\em Three types of solutions at zero temperature.--}  To study the quantum phase transition of the system, we shall focus on solutions at zero temperature. We choose the ansatz of the background to be
\bea
ds^2&=&u(-dt^2+dz^2)+\frac{dr^2}{u}+f(dx^2+dy^2)\,,\nn\\
\label{metric} 
\Phi&=&\phi(r)\,,\\
B_{xy}&=&B(r)\,\nn.
\eea

At the UV boundary, the fields $\phi(r)$ and $B(r)$ behave as
\be
\phi= \frac{M}{r}+\cdots\,,~~~~~ B= b r +\cdots\,,
\ee where $M$ and $b$ correspond to the source for the two dual operators. All through the paper we will work at $b=1$, i.e. all dimensionful parameters are compared to $b$. To solve these equations, we need to identify the near horizon boundary condition. It turns out we have three different kinds of near horizon geometries. With proper irrelevant deformations, these solutions flow to the UV AdS$_5$ with different boundary values of $M/b$. 

 \para
\noindent {\it Topological phase.} The near horizon solution for the topological phase is
\bea
u&=&\frac{1}{8}(11+3\sqrt{13}) r^2\Big(1+\delta u\, r^{\alpha_1} \Big)\,,\nn\\
f&=& \sqrt{\frac{2\sqrt{13}}{3}-2}\, b_0 r^\alpha \Big(1+\delta f\, r^{\alpha_1} \Big)\,,\nn\\
\phi &=& \phi_0 r^{\beta}\,,\nn\\
B&=&b_0 r^\alpha \Big(1+\delta b\, r^{\alpha_1} \Big)\,,\nn
\eea
where $(\alpha, \beta, \alpha_1)=(0.183, 0.290
, 1.273)$, $(\delta f, \delta b)=(-2.616, -0.302)\delta u$ for the parameter values that we have fixed above.  $b_0$ can be set to $1$ by the transformation $(x,y)\to c (x,y)$. Moreover, the near horizon geometry has a Lifshitz symmetry at leading order
\be\label{eq:lif}
(t,z,r^{-1})\to c(t,z,r^{-1})\,,~~~(x,y)\to c^{\alpha/2} (x,y)\,,
\ee
which can be used to set $\delta u =\pm 1$ and $\delta u=-1$ flows the geometry to AdS$_5$. Thus in the IR, we only have a unique free parameter $\phi_0$.
 \para
We integrate this solution to the boundary and for this type of near horizon boundary conditions we can only find solutions for $M/b< 1.717$. When the value of $\phi_0$ is zero, the boundary reaches $M/b=0$. When $\phi_0$ grows, $M/b$ becomes larger and closer to the critical value $1.717$.  To see that this corresponds to the NLSM phase, we will give evidence from fermion spectral functions below. 

\para
\noindent{\it Critical point.} The near horizon solution for the critical point including irrelevant deformations is 
\bea
u&=& u_c r^2 (1+\delta u\, r^{\beta_1})\,,\nn\\
f&=& f_c r^{\alpha_c} (1+\delta f\, r^{\beta_1})\,,\nn\\
\phi &=& \phi_c  (1+\delta \phi\, r^{\beta_1})\,,\nn\\
B&=& b_c r^{\alpha_c} (1+\delta b\, r^{\beta_1})\,,\nn
\eea
with  
\be
(u_c, f_c, \alpha_c, \phi_c)\simeq(3.076, 0.828 b_c, 0.292, 0.894)\,, \nn
\ee and 
\be
\beta_1=1.272\,,~~~~~
(\delta u, \delta f, \delta b)=(1.177, -2.771,-0.409)\delta\phi\,. \nn
\ee 

Using the transformation $(x,y)\to c (x,y)$, we set $b_c=1$. 
Without the deformation, the near horizon has the same type of Lifshitz symmetry (\ref{eq:lif}) with $\alpha$ replaced by $\alpha_c$. 
Utilizing this symmetry, $\delta\phi$ could be chosen to be 
 $\delta\phi=-1$ to flow the geometry to $AdS_5$ at the boundary. There is no free parameter in the IR and the geometry is unique. At the boundary we find the critical $M/b\simeq 1.717$.

\begin{figure}[h!]
\begin{center}
\includegraphics[width=0.43\textwidth]{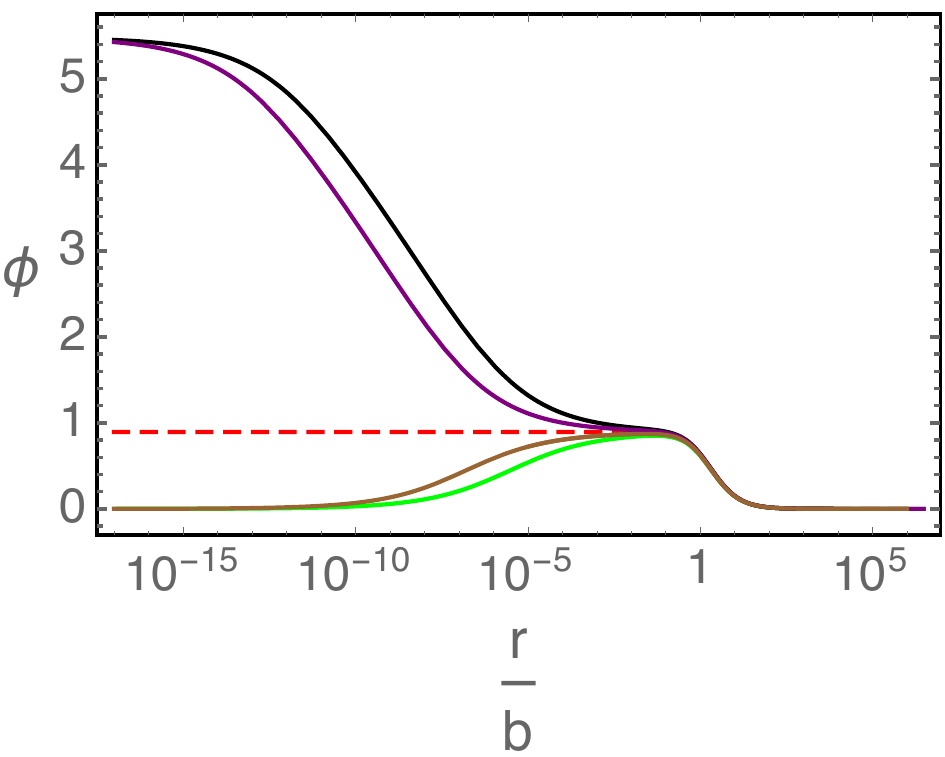}
\includegraphics[width=0.44\textwidth]{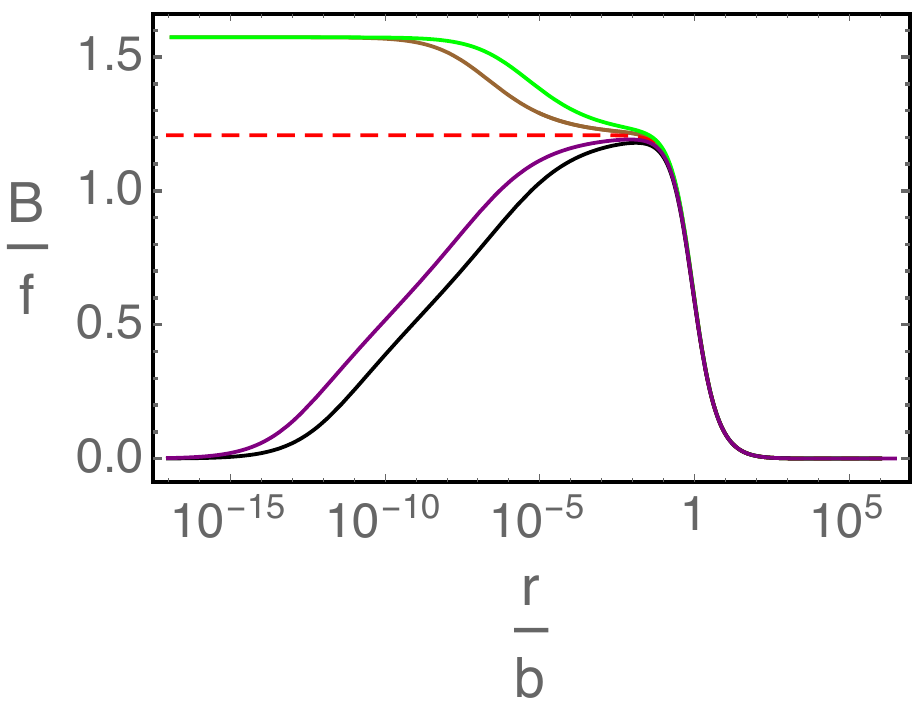}
\end{center}
\vspace{-0.48cm}
\caption{\small The bulk profile for the scalar field $\phi$ and the two form field $B/f$ for different values of $M/b=1.682$ (green), $1.702$ (brown), $1.717$ (red), $1.733$ (purple), $1.750$ (black).}
\label{fig:bg}
\end{figure}

\begin{figure}[h!]
\begin{center}
\includegraphics[width=0.6\textwidth]{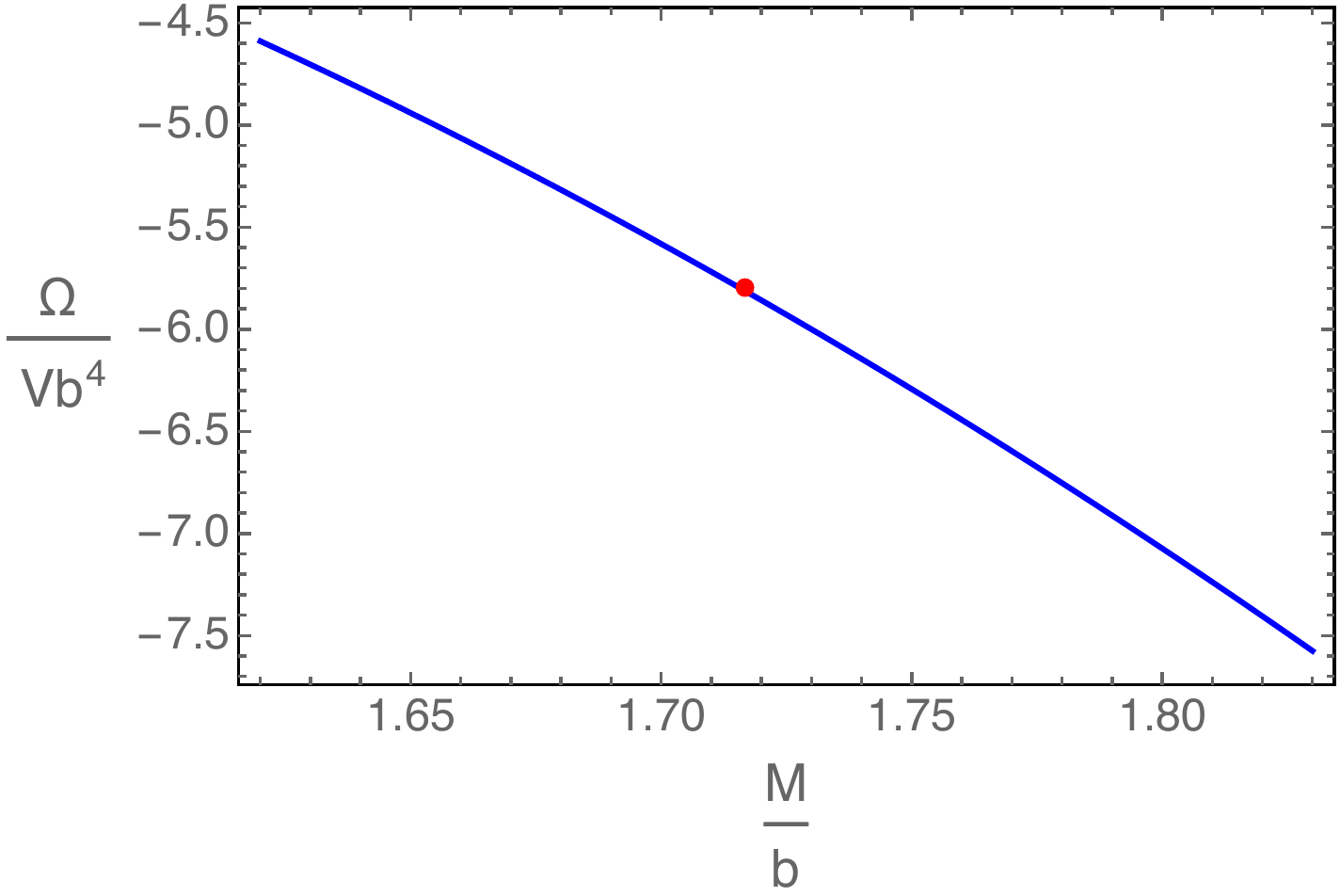}
\end{center}
\vspace{-0.48cm}
\caption{\small The dependance of free energy on $M/b$. The red dot is the value of free energy density at the critical point. Approaching the transition point from both the topological phase and the trivial phase, the system is continuous and smooth.}
\label{fig:fe}
\end{figure}

\para
\noindent{\it Trivial phase.} The near horizon solution for the trivial phase is
\bea
u&=&\big(1+\frac{3}{8\lambda_1}\big)r^2\,,\nn\\
f&=& r^2\,,\nn\\
\phi&=&\sqrt{\frac{3}{\lambda_1}}+\phi_1 r^{\frac{2\sqrt{160\lambda_1^2+84\lambda_1+9}}{3+8\lambda_1}-2}\,,\nn\\
B&=& b_1 r^{2\sqrt{2}\sqrt{\frac{3\lambda+\lambda_1}{3+8\lambda_1}}}\,.\nn
\eea 
Note that the $\phi_1$- and $b_1$-terms are irrelevant deformations that flow the near AdS$_5$ leading order exact solution to 
asymptotic $AdS_5$ solutions. For this type of near horizon boundary conditions we can only find solutions for $M/b>1.717$. For the critical and the trivial solutions here, the scalar field $\phi$ is a finite constant at the horizon and the system is partially gapped. 
 \para
In Fig. \ref{fig:bg}, we show the bulk behavior of $\phi$ and $B/f$ for different values of $M/b$ which correspond to different near horizon geometries. Close to the critical $M/b$, the near horizon solution flows to the critical solution quickly. Fig. \ref{fig:fe} is the free energy of this system, which shows that the system is continuous when crossing the quantum phase transition point. From Figs.  \ref{fig:bg} and \ref{fig:fe} we could see that though the phase transition is a very continuous one, the bulk IR solutions seem to be discontinuous. This is a common feature of many holographic continuous phase transitions, especially BKT phase transitions \cite{Iqbal:2011aj, Donos:2012js,Hartnoll:2011pp, Liu:2013yaa}. Near the transition critical point of such systems, there exists an IR critical scale below which the solutions of the two phases (or the solution of one of the phases when there is only one solution at each region of the phase diagram) start to deviate from the critical solution while above this IR critical scale, the solutions of the two phases (or the solution of one of the phases) converge to the critical solution. This IR scale depends on how far the system is way from the critical point and could be exponentially small near the critical point for very continuous phase transitions, e.g. BKT phase transitions \cite{Iqbal:2011aj}. As the system approaches the transition critical point, this IR scale decreases and becomes infinitely small so that the solutions of the two phases converge to the critical solution for almost all the bulk spacetime.  Thus the free energy also shows a very continuous behavior even though the IR solutions look discontinuous.

 \para
Note that the mechanism for the topological quantum phase transition here is different from the Breitenlohner-Freedman bound mechanism for many holographic quantum phase transitions, including holographic superconductors \cite{Iqbal:2011aj}, metal-insulator phase transitions \cite{Donos:2012js}, etc.. Here the IR scaling dimensions of the fields do not change for each phase and different phases are due to different values of sources at the UV. There is no local order operator associated with this phase transition 
similar to the holographic quantum phase transitions in \cite{Hartnoll:2011pp}.

\section{Fermion spectral functions}
\label{sec3}

\begin{figure}[h!]
\begin{center}
\includegraphics[width=0.56\textwidth]{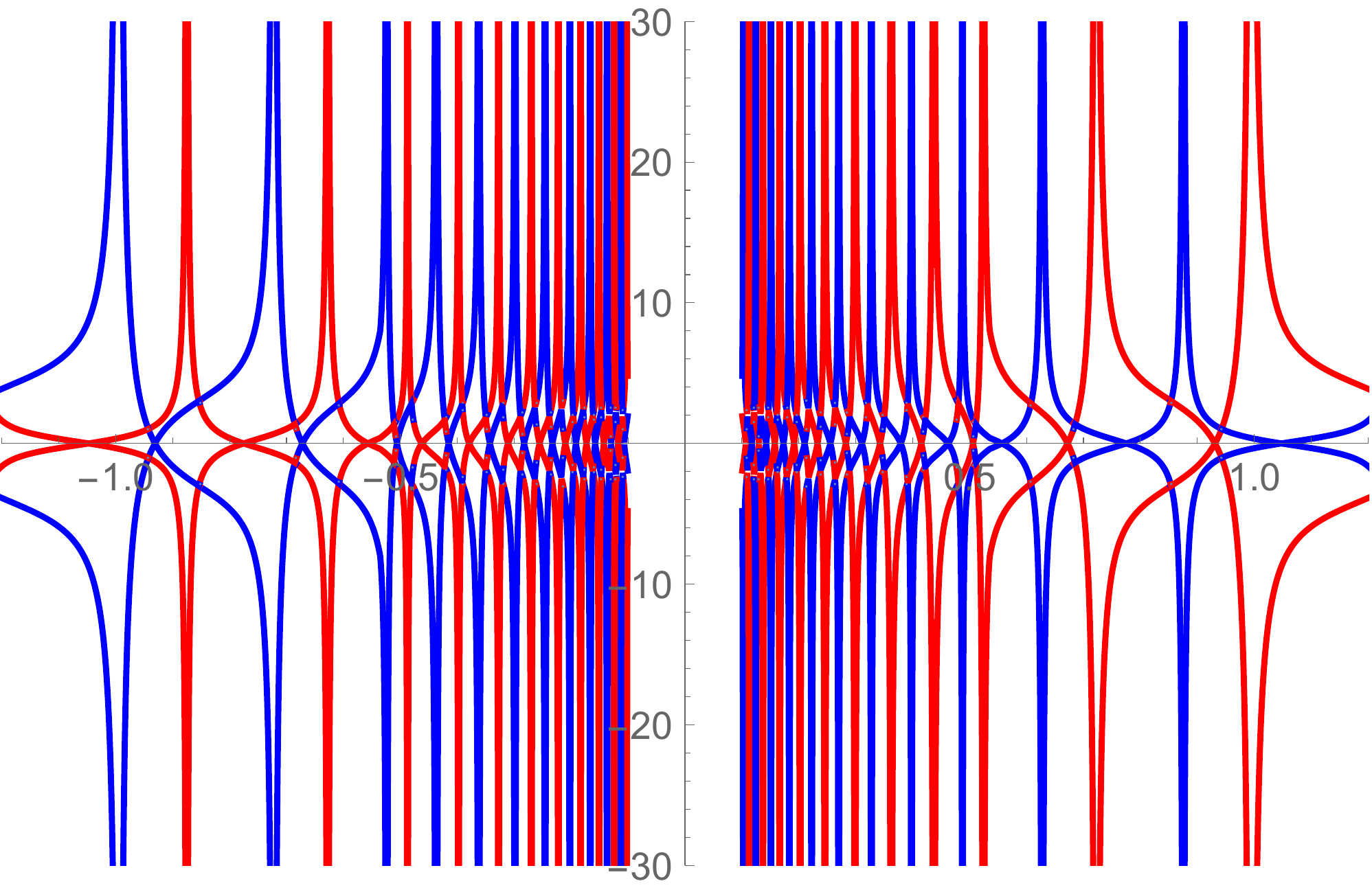}
\end{center}
\vspace{-0.2cm}
\caption{\small Eigenvalues of $-G^{-1}(0, k)$ for $M/b\simeq 0.0013$ as a function of $k_x$. Red and blue curves represent two different sets of eigenvalues. This corresponds to an effective band structure of the topological Hamiltonian of this system, which captures the same topological structure of the original system. This figure also agrees qualitatively with the density plot of spectral densities in the $\omega$-$k_x$ plane of the original system. Figure taken from \cite{Liu:2018djq}.
}
\label{fig:spec}
\end{figure}

For NLSMs there is no smoking gun transport coefficient like anomalous Hall conductivity for WSMs. An immediate and straightforward test would be to see if there is indeed a closed nodal loop from the dual fermion spectral functions. 
We will show that the fermion spectral function in this system confirms that in the nodal line semimetal phase, there is indeed a peak\footnote{We calculate the dual retarded Green's function from bulk probe fermions by numerics and keep a very small $\omega=10^{-5}$ for numerical convenience.  This peak becomes a pole when the small value of $\omega$ is removed. } at $k_x^2+k_y^2=k_0^2$ where the value of $k_0$ decreases with $M/b$ increasing and finally becomes zero at the critical point. We expect that we could also see the nontrivial topology in the structure of the fermion spectral functions, which we hope to report in the future work. 
 \para
The prescription for calculating the holographic fermion spectral function could be found in \cite{Iqbal:2009fd, Liu:2009dm, Cubrovic:2009ye}. In five dimensions, one bulk Dirac spinor corresponds to one chirality of boundary Weyl spinor.  We utilize two spinors with  
one spinor to use standard quantization to calculate the dual fermion spectral function and change the sign of Fermi mass $m$ and use alternative quantization to calculated the fermion spectral function of the other chirality.
since in both the holographic Weyl/nodal line semimetal model, the two chiralities interact with each other due to the axial symmetry breaking terms. As can be seen from formula (\ref{conservation2}) both the mass term and the $b_{\mu\nu}$ term break the axial symmetry and couple spinors of two chiralities together. The action of the two spinors $\Psi_1$ and $\Psi_2$ are
\be
S_\text{fermion}=S_1+S_2+S_{\text{int}}\,,
\ee 
where
\bea
S_1&=&\int d^5x \sqrt{-g} i\bar{\Psi}_1\big(\Gamma^a D_a -m_f \big)\Psi_1\,,\nonumber \\
S_2&=&\int d^5x \sqrt{-g} i\bar{\Psi}_2\big(\Gamma^a D_a +m_f \big)\Psi_2\,,\nonumber \\
S_{\text{int}}&=&-\int d^5x \sqrt{-g}\big( i\eta_1\Phi\bar{\Psi}_1 \Psi_2+i\eta_1^* \Phi^*\bar{\Psi}_2 \Psi_1\big)+S_B\,,\nn
\eea and
\be\label{sb}
S_B=\int d^5x \sqrt{-g} i(\eta_2 B_{ab}\bar{\Psi}_1 \Gamma^{ab}\gamma^5\Psi_2-\eta_{2}^*B_{ab}^*\bar{\Psi}_2\Gamma^{ab}\gamma^5\Psi_1)\,.\ee
\para 
Note that the Lorentz invariance in the tangent space has been explicitly broken in the bulk and this is the exact choice to produce the specific coupling for the fermions to generate a nodal line in the spectrum. Here
\be
D_a=\partial_a-\frac{i}{4}\omega_{\underline{m}\underline{n},a}\Gamma^{\underline{m}\underline{n}}-iq_3 A_a
\ee
where $\omega_{\underline{m}\underline{n},a}$ is the bulk spin connection. Note that $a$ and $\underline{m}$ are the bulk spacetime index and the tangent space index respectively. $\bar{\Psi}=\Psi^\dagger \Gamma^{\underline{t}}$,
$\Gamma^{ab}=e_{\underline{m}a}e_{\underline{n}b}\frac{i}{2}[\Gamma^{\underline{m}},\Gamma^{\underline{n}}]$. $m_f$ is the mass of the bulk spinor which determines the scaling dimension of the dual Fermionic operator and we choose $m_f=-1/4$ so that there could be poles in the dual fermion spectral functions \cite{Iqbal:2009fd}. We also set $\eta_1=\eta_2=1$. 
The equations of motion are
\bea
\big(\Gamma^a D_a -m_f \big)\Psi_1-\big(\eta_1 \phi-\eta_2 B_{ab}\Gamma^{ab} \gamma^5\big)\Psi_2=0\,,\nonumber\\
\big(\Gamma^a D_a +m_f \big)\Psi_2-\big(\eta_1 \phi+\eta_2^* B_{ab}^*\Gamma^{ab} \gamma^5 \big)\Psi_1=0\,,
\eea 
where  
the five dimensional $\Gamma$ functions are $
\Gamma^{\underline{\mu}}=\gamma^\mu$, $ \Gamma^{\underline{r}}=\gamma^5.$ We expand the fermion field as $$\Psi_l= (uf)^{-1/2} \psi_l e^{-i\omega t+i k_x x+i k_y y+i k_z z}$$ with $l=(1,2)$ and the corresponding Dirac equation can be written as 
\bea
&&\Bigg(\Gamma^{\underline{r}}\partial_r+\frac{1}{u}\Big(-i\omega \Gamma^{\underline{t}}+ik_z \Gamma^{\underline{z}}\Big)+
\frac{1}{\sqrt{uf}}\Big(ik_x\Gamma^{\underline{x}}+ik_y\Gamma^{\underline{y}}\Big)+(-1)^l\frac{m_f}{\sqrt{u}}\Bigg)\psi_{l}
\nn\\&&~~~~~-\Bigg(\eta_1\frac{\phi}{\sqrt{u}}+(-1)^l\eta_2\frac{b}{\sqrt{u}f}\Gamma^{\underline{x}\underline{y}}\gamma^5\Bigg)\psi_{3-l}=0\,.
\eea
 \para
We can solve this as a coupled system of 8 functions. At the horizon the ingoing boundary condition depends on the near horizon geometry of each phase. For the topologically trivial phase, the near horizon ingoing solution for nonzero $k$ while $\omega\to 0$ is real just as the pure AdS$_5$ case in \cite{Iqbal:2009fd}. Thus the imaginary part of the Green's function is automatically zero for nonzero $k$ and the Fermi momentum should stay at $k=0$. For the topologically nontrivial and critical phases, the near horizon ingoing boundary condition is
\be
\psi_{l}\simeq
  e^{\frac{i\sqrt{w^2-k_z^2}}{u_0r}}
\begin{pmatrix} 
z^{l}_1 (1+\cdots) \\
\vspace{-.3cm}\\
z^{l}_ 2 (1+\cdots)  \\
\vspace{-.3cm}\\
i\frac{\sqrt{\omega^2-k_z^2}}{\omega-k_z}z^{l}_1 (1+\cdots)   \\
\vspace{-.3cm}\\
i\frac{\sqrt{\omega^2-k_z^2}}{\omega+k_z}z^{l}_2  (1+\cdots)  \end{pmatrix}
\ee
with $l=(1,2)$ for $w>k_z$, where $``\cdots"$ denote subleading terms. This is because the terms proportional to $k_x$ and $k_y$ are not so important compared to terms with $\omega$ and $k_z$ as $g^{xx}$ is not as divergent as $g^{zz}$ and $g^{tt}$ at the horizon. This also means that for nonzero $k_z$ and $\omega\to 0$, the retarded Green's function is real and the Fermi surface could only stay at $k_z=0$, which is consistent with the fact that this corresponds to a nodal line semimetal in the $x$-$y$ plane.
 \para
Near the boundary $r\to\infty$,
\be
\psi_{l}=\begin{pmatrix} 
a^{l}_1  &&r^{m_f}+\cdots \\
\vspace{-.3cm}\\
a^{l}_ 2  &&r^{m_f}+\cdots \\
\vspace{-.3cm}\\
a^{l}_ 3 && r^{-m_f}+\cdots \\
\vspace{-.3cm}\\
a^{l}_ 4 && r^{-m_f}+\cdots  \end{pmatrix}\,,
\ee 
with $l=(1,2)$.
 \para
Because the two chiralities couple to each other, $\psi_{1,2}$ will also source expectation values of $\psi_{2,1}$ respectively. To calculate the retarded Green's function, we need four different horizon boundary conditions and get four sets of source and expectation values. We denote the four boundary conditions as I, II, III, IV respectively and the source and expectation matrix would look like

$$~~~~~M_s=\begin{pmatrix} 
a^{1,I}_1&~&a^{1,II}_1&~&a^{1,III}_1&~&a^{1,IV}_1\\
\vspace{-.3cm}\\
a^{1,I}_2&~&a^{1,II}_2&~&a^{1,III}_2&~&a^{1,IV}_2\\
\vspace{-.3cm}\\
a^{2,I}_3&~&a^{2,II}_3&~&a^{2,III}_3&~&a^{2,IV}_3\\
\vspace{-.3cm}\\
a^{2,I}_4&~&a^{2,II}_4&~&a^{2,III}_4&~&a^{2,IV}_4\\
\end{pmatrix} 
~~~~~\text{and} ~~~~ M_e=\begin{pmatrix} 
-a^{2,I}_1&~&-a^{2,II}_1&~&-a^{2,III}_1&~&-a^{2,IV}_1\\
\vspace{-.3cm}\\
-a^{2,I}_2&~&-a^{2,II}_2&~&-a^{2,III}_2&~&-a^{2,IV}_2\\
\vspace{-.3cm}\\
a^{1,I}_3 &~&a^{1,II}_3&~&a^{1,III}_3&~&a^{1,IV}_3\\
\vspace{-.3cm}\\
a^{1,I}_4&~&a^{1,II}_4&~&a^{1,III}_4&~&a^{1,IV}_4\\
\end{pmatrix}\,.$$
 \para
The Green's function is determined by $G=- M_eM_s^{-1}\Gamma^t$. After getting $G$ we find eigenvalues of $G$ and read the imaginary part of the four eigenvalues. With the above ingredients one can compute the spectral function. A plot for the spectral function $G^{-1}(0, k_x)$ at $k_z=\omega=0$ for a finite regime of $k_x$ is shown in Fig. \ref{fig:spec}, which is taken from \cite{Liu:2018djq}. In the framework of topological systems, we can treat $-G^{-1}(0, k)$ as a topological Hamiltonian \cite{wang-prx, interaction1}  which essentially determines the topological behavior of the system and the eigenvalues plot would agree qualitatively with the spectral density plot in the $\omega$-$k_x$ plane. Different from the weakly coupled band structure in Fig. \ref{fig:phase} the strong interaction hybridize all the four bands to have multiple poles. We use different colors to distinguish different bands in Fig. \ref{fig:phase}. More explanations and details on the fermion spectral functions of this system and topological invariants could be found in the follow-up work \cite{Liu:2018djq}, which are based on the construction in this paper.

\begin{figure}[h!]
\begin{center}
\includegraphics[width=0.4\textwidth]{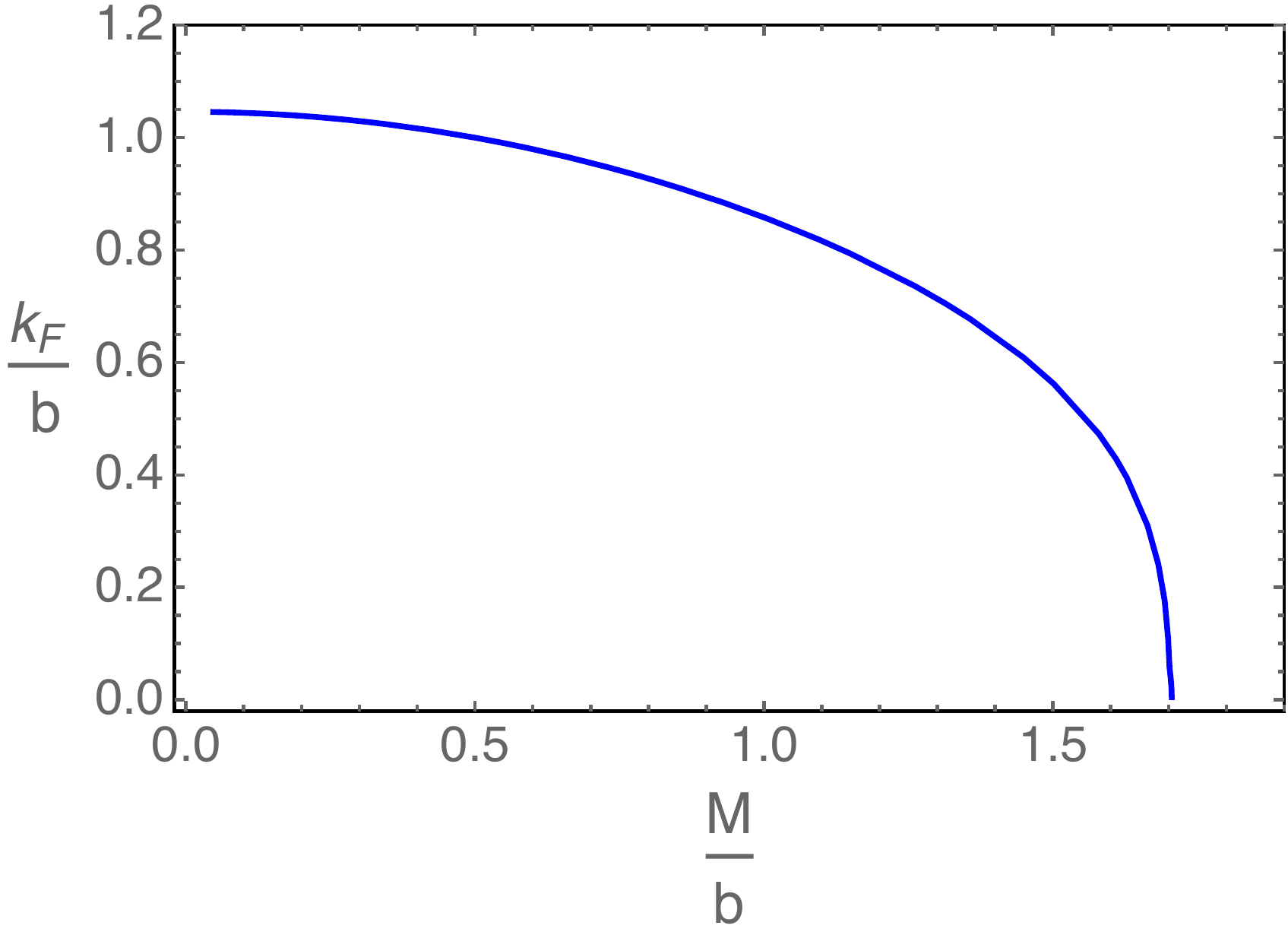}
\includegraphics[width=0.43\textwidth]{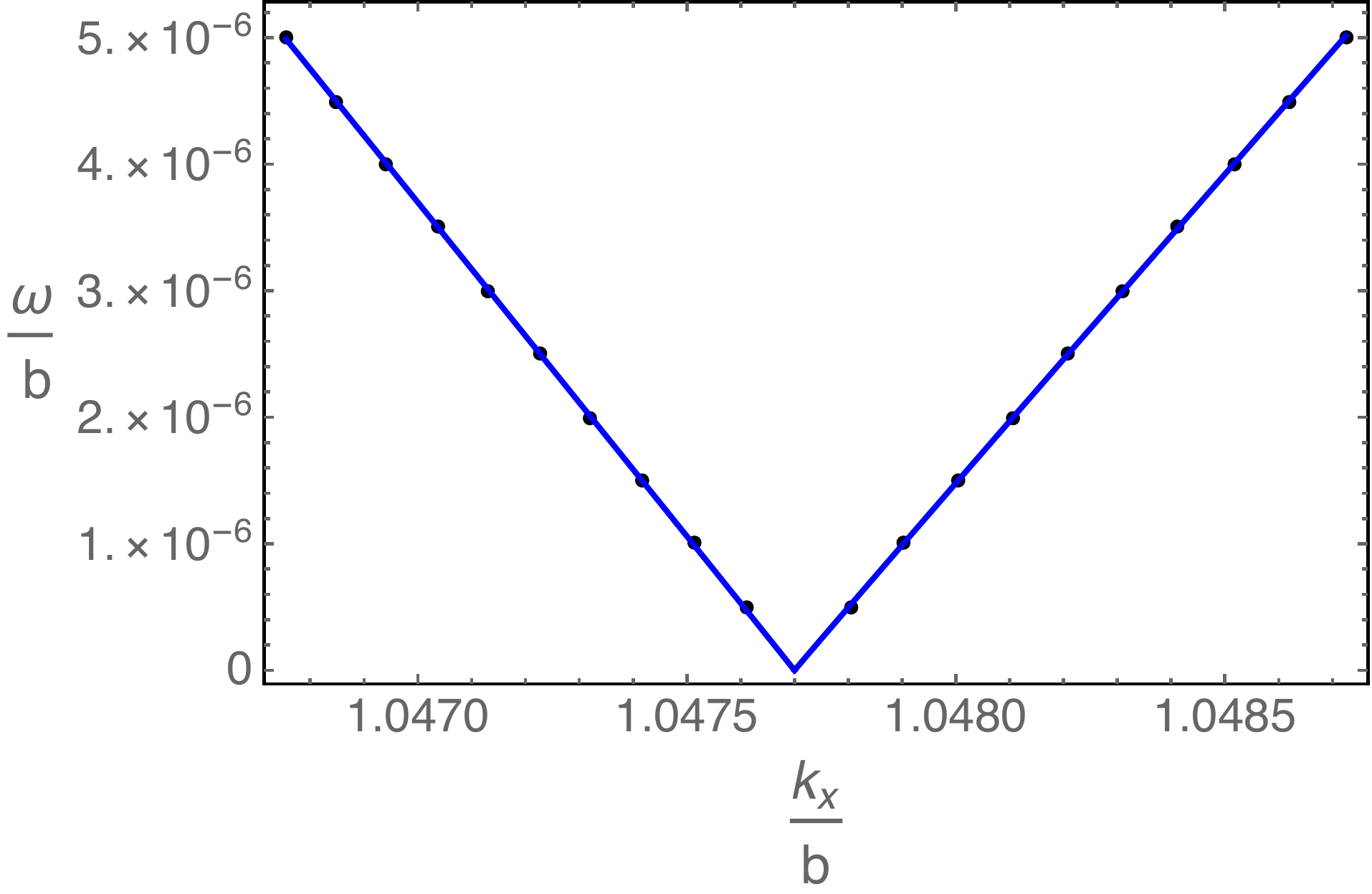}
\end{center}
\vspace{-0.4cm}
\caption{\small {\em Left:} The Femi momentum $k_F=\sqrt{k_x^2+k_y^2}$ in the holographic NLSMs for $k_z=0$. In the critical point or the trivial phase, there is no Fermi surface at finite $k$ and the pole is located at $k=\omega=0$. {\em Right:} An example of dispersion relation. For $M/b\simeq 0.0013$, the best fitting curves for the poles are almost linear in $k_F-k_x$, i.e. for $k_x<k_F$, $\omega\simeq 0.005 (1.0477-k_x)^{0.998}$; while for $k_x>k_F$, $\omega\simeq 0.005 (k_x-1.0477)^{0.994}$. }
\label{fig:fs}
\end{figure}

\para
In the following we summarise the properties for the Green's  function in the three phases.
\begin{itemize}
\item In all the three phases, when $k_z$ is not zero, the zero frequency retarded Green's function is real and does not have an imaginary part.  

\item In the topologically trivial phase, the near horizon geometry guarantees that for all values of $k_x, k_y, k_z\neq 0$,  the retarded Green's function is real. The pole stays at $k_x=k_y=k_z=0$ as this is 
a partially gapped trivial semimetal phase. 

\item For the critical solution, we find that two branches of eigenvalues have peaks in the imaginary part at $k_x=k_y=0$ and the other two are still small for all $k_x,k_y$.

\item Different from the weakly coupled NLSMs, for the holographic NLSM phase, the system has multiple and discrete Fermi surfaces at a set of values of $k_F^{i}=\sqrt{k_x^2+k_y^2}$ and $k_z=0$, $\omega\to 0$, which are closed nodal loops in the $k_x$-$k_y$ plane. This feature reflects the 
strong coupling effect of 
the holographic system. At each nodal line momentum, there is a sharp peak (which becomes a pole at $\omega=0$) in the imaginary part of two eigenvalues of the Green's function 
and the imaginary part of the other two eigenvalues are very small indicating that the other two are gapped at this momentum. This different behavior in the $k_z$ and $k_x$, $k_y$ directions are partially caused by the IR Lifshitz geometry. 

\item Each pair of adjacent nodal lines come from opposite sets of two bands \cite{Liu:2018djq}. The interval between two adjacent nodal lines gets larger as $k_F$ increases. The position of $k_F$ of each branch of nodal line decreases as $M/b$ increases and reaches zero at the quantum critical point. The left plot in Fig. \ref{fig:fs} shows the value of one branch of $k_F^{i}$ as a function of $M/b$. For each branch of nodal lines, we find that the dispersion in both the $k_z$ and $k_x$ directions are almost linear. The right plot in Fig. \ref{fig:fs} shows the dispersion in the $k_x$ direction of one branch of nodal lines at $M/b\simeq 0.0013$. A figure taken from \cite{Liu:2018djq} showing the fermion spectral structure could be found in Fig. \ref{fig:spec}. More details about the fermion spectral functions and topological invariants of this system could be found in \cite{Liu:2018djq}. 

\end{itemize}
 \para
The spectral function behavior confirms that the topological phase corresponds to a NLSM with multiple nodal lines and the critical and trivial solutions correspond to trivial semimetals. Strong coupling effect produces more complicated topological structures in the NLSM phase. Nontrivial topological invariants could be obtained in the holographic NLSM phase from the dual Green functions \cite{Liu:2018djq} using the topological Hamiltonian method developed in \cite{wang-prx}. Moreover, in the holographic NLSM phase as the subleading order of $\phi$ grows bigger the radius of the nodal line circle becomes smaller. This confirms that small perturbations would not gap the system while only making the Fermi nodal circle bigger or smaller.

\section{Topological structure in the solution space}
\label{sec4}

In the bulk there is an intrinsic topological structure for this system 
from the near horizon configurations of the matter fields $B_{ab}$ and $\Phi$.
 \para
The near horizon leading order behavior of $B$ and $\phi$ are determined by their IR conformal dimensions $\delta_{\pm}^{B,\phi}$ in the backreacted geometry. As $ r^{-\delta_{+}^{B,\phi}}$ is too divergent in the IR, we have  $B,\phi\sim c_{B,\phi}r^{-\delta_{-}^{B,\phi}}$. One crucial observation is that with the interaction term between $B$ and $\phi$, $c_{B}$ and $c_\phi$ cannot both be nonzero simultaneously. This divided the solutions into three categories: $c_{\phi}=0$ while $c_B\neq 0$ corresponding to the topologically nontrivial semimetal solution, $c_{B}=0$ while $c_{\phi}\neq 0$ corresponding to the trivial phase and $c_{\phi}=c_{B}=0$ corresponding to the critical phase where near horizon solutions are subleading terms. In other words, the interaction term always changes the IR scaling dimension of at least one of the fields and it is not possible to keep both scaling dimensions unchanged. Note that this structure is still true with the presence of $\lambda_1\phi^4$ term where $\phi$ is at most $O(1)$ at leading order and we can substitute  $c_{\phi} r^{-\delta_{-}^{\phi}}$ with $c_{\phi}$ in this case. 
 \para
This structure indicates that when we have a NLSM solution, we cannot find a small perturbation in the IR with $\phi$ at  $\phi\sim c_\phi r^{-\delta_{-}^{\phi}}$ or $c_\phi$ in the case with $\lambda_1\phi^4$ term, which could (partially) gap the semimetal. This means that the interaction between $B$ and $\phi$ at IR is the intrinsic reason that the nodal line solution is topological. The topological NLSM phase could only become partially gapped after a quantum phase transition passing through the critical point where the nodal line becomes a nodal point.
\subsection{A general framework} This holographic NLSM shares the same mathematical structure as the holographic WSM \cite{Landsteiner:2015pdh}. We propose that there exists a general framework in holography to produce strongly coupled topological gapless states and the key ingredients of the holographic topological structure are:
\vspace{-.1cm}
\begin{itemize}
\item 
A holographic system with at least two interacting fields, with one corresponding to a mass operator, and the other corresponding to an operator that deforms the topology of the Fermi surface, which we denote as $\phi$ and $A$ for illustration. 
The IR interaction between the two operators produces the topological structure of the solution space in the bulk.
\item 
At the horizon there are three types of solutions: $A$ (or $\phi$) is nonzero at leading order with $r^{-\delta_-^{A,\phi}}$ while $\phi$ (or $A$) is at subleading order sourced by $A$ (or $\phi$) and a critical solution with two fields both at subleading order sourcing each other. The fact that $A$ and $\phi$ cannot both be at leading order with $r^{-\delta_-^{A,\phi}}$ at the horizon implies that the semimetal phase cannot be gapped by small perturbations.
\end{itemize}

We expect that a general holographic topological semimetal state shares the properties above. It is expected to be able to describe topological gapped states of matter after introducing bulk fields dual to boundary gapped operators. As there is an intrinsic bulk topological structure for the system that does not require any knowledge of the dual fermion spectral functions, it is possible that this framework would shed light on a better understanding of the deeper organizational principles of strongly coupled topological gapless states of matter and predict new kinds of strongly coupled topological semimetal states. 

\vspace{.2cm}
\begin{acknowledgments}
 We would like to thank Rong-Gen Cai, Chen Fang, Carlos Hoyos, Ling-Yan Hung, Elias Kiritsis, Karl Landsteiner, Shunqing Shen,  Gang Su, Zhong Wang, Hong Yao, Hui Zhai and Fuchun Zhang for useful discussions.  This work is supported by the National Key R\&D Program of China (Grant No. 2018FYA0305800) and by the Thousand Young Talents Program of China. The work of Y.L. was also supported by the NFSC Grant No.11875083 and a grant from Beihang University. The work of Y.W.S. has also been partly supported by starting grants from UCAS and CAS, and by the Key Research Program of the Chinese Academy of Sciences (Grant No. XDPB08-1), the Strategic Priority Research Program of Chinese Academy of Sciences, Grant No. XDB28000000.
\end{acknowledgments}

\appendix

\section{Equations of motion}
\label{secaa}

The equations of motion for the action in the main text are
\bea\label{eq:ein1}
R_{ab}-\frac{1}{2}g_{ab}(R+12)-T_{ab}&=&0\,,\\
\nabla_b \mathcal{F}^{ba}+2\alpha \epsilon^{abcde} F_{bc}\mathcal{F}_{de}&=&0\,,\\
\nabla_b F^{ba}+\alpha \epsilon^{abcde} (F_{bc}F_{de}+\mathcal{F}_{bc}\mathcal{F}_{de})
~~~~~~&&\nn\\
-iq_1\big(\Phi^*D^a\Phi-(D^a\Phi)^*\Phi\big)
-\frac{iq_2}{\eta}\big(B_{ca}^*\mathcal{D}^{[b}B^{ca]}-(\mathcal{D}^{[b}B^{ca]})^*B_{ca}\big)&=&0\,,\\
D_a D^a\Phi-\partial_{\Phi^*} V_1-\lambda\Phi B_{ab}^*B^{ab}&=&0\,,\\
\frac{1}{\eta}\mathcal{D}^a \mathcal{D}_{[a} B_{bc]}-m_2^2 B_{ab}-\lambda \Phi^*\Phi B_{ab}&=&0\,,
\eea
where 
\bea
T_{ab}&=&\frac{1}{2}\Big[\mathcal{F}_{ac}\mathcal{F}_{b}^{~c}-\frac{1}{4}g_{ab}\mathcal{F}^2\Big]+
\frac{1}{2}\Big[F_{ac} F_{b}^{~c}-\frac{1}{4}g_{ab}F^2\Big]+\frac{1}{2}\big((D_a\Phi)^*D_b\Phi+(D_b\Phi)^*D_a\Phi\big)\nn\\
&&+(m_2^2+\lambda|\Phi|^2)(B_{ac}^*B_b^{~c}+B_{bc}^*B_a^{~c})+\frac{1}{2\eta}\big((\mathcal{D}_{[a}B_{cd]})^*\mathcal{D}_{[b}B^{cd]}+(\mathcal{D}_{[b}B_{cd]})^*\mathcal{D}_{[a}B^{cd]}\big)\nn\\
&&-\frac{1}{6\eta}(\mathcal{D}_{[m}B_{cd]})^*(\mathcal{D}_{[m}B^{cd]})g_{ab}-\frac{1}{2}\Big((D_c \Phi)^*(D^c\Phi)+V_1+V_2+\lambda|\Phi|^2B_{cd}^*B^{cd} \Big)g_{ab}\,.\nn
\eea

 \para
In this system, the scalar and the two form fields both have sources at the boundary, therefore for the system to be normalizable at the boundary and without loss of generality we choose the conformal dimension for the source of the dual mass term and the $\bar{\psi}\gamma^{\mu\nu}\psi$ term in (\ref{eq:1}) to be 1. Thus the massive scalar and 2-form fields in the bulk have the values for the mass to be $m_1^2=-3$ and $m_2^2=1$. Furthermore, for simplicity, we set $q_1=q_2=1$, $\lambda=1$, $\lambda_1=0.1$ and $\eta=1.$
 \para
After a variation of the total action with respect to the gauge fields, we can obtain the dual consistent currents and they satisfy
\bea
\partial_\mu J^\mu_{\text{{con}}}&=&0\,,\nn\\
\partial_\mu J^\mu_{5\text{{con}}}&=&\lim_{r\to\infty}\sqrt{-g}\bigg(-\frac{\alpha}{3}\epsilon^{r\alpha\beta\rho\sigma}(F_{\alpha\beta}F_{\rho\sigma}+\mathcal{F}_{\alpha\beta}\mathcal{F}_{\rho\sigma})+iq_1\Big[\Phi^*(D^r\Phi)-\Phi(D^r\Phi)^*\Big]+\nn\\
&&~~~~~~~~~+\frac{iq_2}{\eta}\big(B_{\mu\nu}^*\mathcal{D}^{[r}B^{\mu\nu]}-(\mathcal{D}^{[r}B^{\mu\nu]})^*B_{\mu\nu}\big)\bigg)+\text{c.t.}\,.\nn
\eea
Here we have not explicitly shown the counterterm for simplicity and the above conservation can be further simplified in the radial gauge. The point is that the last two terms contribute only when the non-normalisable mode of the scalar filed or two-from field is switched on and it is straightforward to see that the above identities are of the same structure of the weakly coupled theory (\ref{conservation}, \ref{conservation2}). Thus this holographic model is expected to go beyond the weakly coupled theory to a strongly coupled nodal line semimetal model.

\subsection{Zero temperature}

With the ansatz of the zero temperature solution in the main text the corresponding equations of motion are 
\bea
\frac{f''}{f}-\frac{u''}{u}+\frac{f'u'}{2f u}-\frac{ u'^2}{2u^2}+\frac{4}{f^2}\bigg(\frac{B'^2}{\eta}+\frac{\lambda B^2\phi^2}{u}+\frac{m_2^2 B^2}{u}\bigg)&=&0\,,\nn\\
\frac{\phi'^2}{2}+\frac{6}{u}-\frac{u'}{ u}\bigg(\frac{f'}{f}+\frac{u'}{4u}\bigg)+\frac{B'^2}{\eta f^2}-\frac{\phi^2}{2u}\bigg(m_1^2+\frac{\lambda_1}{2}\phi^2\bigg)
-\frac{f'^2}{4f^2}-\frac{B^2}{uf^2}\bigg(m_2^2+\lambda\phi^2\bigg)&=&0\,,\nn\\
\label{eomphi}
\phi''+\left(\frac{3u'}{2u}+\frac{f'}{f}\right)\phi'
   -\big(m_1^2+\lambda_1 \phi^2+\frac{2\lambda B^2}{f^2}\big)\frac{\phi}{u}&=&0\,,\nn\\
   \label{eomB}
\frac{B''}{\eta}+\frac{B'}{\eta}\bigg(\frac{3u'}{2u}-\frac{f'}{f}\bigg)-\frac{B}{u}\big(m_2^2+\lambda\phi^2\big)&=&0\,.\nn
\eea

 \para
 Close to the boundary $r\to\infty$, we have the following boundary behavior of the fields 
\bea
u&=&r^2-2 b^2-\frac{M^2}{3}+\Big(\frac{4b^4+2\lambda b^2M^2+M^4}{9}+\frac{\lambda_1 M^4}{6}\Big)\frac{\ln r}{r^2}+\frac{u_2}{r^2}+\cdots\,,\\
f&=&r^2-\frac{M^2}{3}+\Big(\frac{4b^4+2\lambda b^2M^2+M^4}{9}+\frac{\lambda_1 M^4}{6}\Big)\frac{\ln r}{r^2}+\frac{f_2}{r^2}+\cdots \,,\\
\phi&=& \frac{M}{r}+\Big(-\frac{M^3}{3}-\frac{\lambda_1 M^3}{2}-b^2M\lambda\Big) \frac{\ln r}{r^3}+\frac{\phi_2}{r^3}+\cdots\,,\\
B&=& b r +\Big(2 b^3-\frac{\lambda b M^2}{2}\Big)\frac{\ln r}{r}+\frac{b_2}{r}+\cdots\,,
\eea
with $f_2=\frac{1}{144}\big(56b^4+48bb_2+14M^4-72 u_2+4b^2M^2 (4+7\lambda) +9M^4 \lambda_1-72 M\phi_2\big)$.
Note that we have set the non-physical free parameter related to the shift symmetry $r\to r+c$ in the bulk to be zero in the above expansion. 
When we extract the boundary data, we should carefully deal with this shift constant. 
 \para
Radially conserved quantity $\partial_r J^r=0$ with $J^r=\sqrt{u}u'f-u^{3/2}f'-\frac{4u^{3/2}}{f}BB'$. From the near boundary behavior, we have\footnote{With the near horizon conditions, we have $f_2-u_2+\frac{\lambda}{2}M^2b^2=0$ which can be used to check the numerical code.}
$J^r=4f_2-4u_2+2\lambda M^2b^2.$ The following scaling symmetry is useful for rescaling the boundary to be asymptotic to standard AdS$_5$ and $b=1.$\\
(1) $(x,y)\to a(x,y)\,, ~~(f, B_{xy})\to a^{-2} (f, B_{xy})$\,;\\
(2) $r\to a r\,, ~~(t,x,y,z)\to a(t,x,y,z)\,, ~~(u,f,B_{xy})\to a^{-2} (u,f,B_{xy})$\,.

\subsection{Finite temperature}

We start from the most general ansatz for the background solutions at finite temperature that are allowed by the symmetry
\bea\label{eq:ansfiniteT}
ds^2&=&-udt^2+\frac{dr^2}{u}+f(dx^2+dy^2)+h dz^2\,\nn\\
\label{metric-finiteT} 
\Phi&=&\phi(r)\,,\\
B_{xy}&=&B(r)\,\nn.
\eea
The equations of motion are
\bea
\frac{f''}{f}-\frac{u''}{u}+\frac{f'h'}{2f h}-\frac{h' u'}{2h u}+\frac{4}{f^2}\bigg(\frac{B'^2}{\eta}+\frac{\lambda B^2\phi^2}{u}+\frac{m_2^2B^2}{u}\bigg)&=&0\,,\\
\frac{f''}{f}+\frac{f'}{f}\left(\frac{h'}{6h}+\frac{2u'}{3u}\right)-\frac{f'^2}{6f^2}-\frac{4}{u}+\frac{ 2B^2}{u f^2}\bigg(m_2^2+\lambda \phi^2\bigg)+~~~~&&\nn\\
+\frac{\phi^2}{3u}\bigg(m_1^2+\frac{\lambda_1\phi^2}{2}\bigg)+\frac{2B'^2}{\eta f^2}-\frac{f'^2}{6f^2}-\frac{h'u'}{6hu}
+\frac{\phi'^2}{3}&=&0\,,\\
\frac{\phi'^2}{2}+\frac{6}{u}-\frac{u'}{2 u}\bigg(\frac{f'}{f}+\frac{h'}{2h}\bigg)-\frac{f'h'}{2fh}+\frac{B'^2}{\eta f^2}-\frac{\phi^2}{2u}\bigg(m_1^2+\frac{\lambda_1}{2}\phi^2\bigg)~~~~\nn\\
-\frac{f'^2}{4f^2}-\frac{B^2}{uf^2}\bigg(m_2^2+\lambda\phi^2\bigg)&=&0\,,\\
\phi''+\left(\frac{f'}{f}+\frac{h'}{2h}+\frac{u'}{u}\right)\phi'
   -\bigg(m_1^2+\frac{2\lambda B^2}{f^2}+\lambda_1 \phi^2\bigg)\frac{\phi}{u}&=&0\,,\\
\frac{B''}{\eta}+\frac{B'}{\eta}\bigg(\frac{u'}{u}+\frac{h'}{2h}-\frac{f'}{f}\bigg)-\frac{B}{u}\big(m_2^2+\lambda\phi^2\big)&=&0\,.
\eea
 \para
Note that the ansatz (\ref{eq:ansfiniteT}) is invariant under two independant transformations 
$$(t,x,y,z,r)\to (a^{-2\alpha-\beta}t,a^\alpha x,a^\alpha y,a^\beta z, a^{2\alpha+\beta} r), ~~(u, f, h, \phi, B)\to (a^{4\alpha+2\beta} u, a^{-2\alpha} f, a^{-2\beta} h, \phi, a^{-2 \alpha} B)  $$ 
with constant $a$ and arbitrary $\alpha$ and $\beta$. The corresponding two Noether currents (radially conserved quantities) are $J_1^r=fh^{3/2}\big(u/h\big)'$ and $J_2^r=\frac{fuh'}{\sqrt{h}}-u\sqrt{h}f'-\frac{4u\sqrt{h}BB'}{f}$. From the near horizon behavior of the fields one concludes that $J_1^r=Ts$ with $s$ the entropy of the system. At zero temperature, we have $J_1^r =0$ which leads to $h=u$ and the ansatz (\ref{eq:ansfiniteT}) is reduced to the zero temperature ansatz in the main text and this confirms that the ansatz made for the zero temperator background is the most generic one. The sencond Noether current $J_2^r$ reduced to the one discussed in the subsection for zero temperature. 

\section{Free energy}
\label{secfe}

For each value of $M/b$ from the boundary, there is only one bulk solution, nevertheless we can compute the free energy of this system to see whether the phase transition is a continuous one. To compute the free energy, we need to be careful with the boundary counterterms.  The total action is 
\be 
S_\text{ren}=S+S_\text{GH}+S_\text{c.t.}
\ee
with the Gibbons-Hawking boundary term $S_\text{GH}=\int_{r=r_\infty} d^4 x\sqrt{-\gamma}(2K)$
and the counterterms
\bea
S_\text{c.t.}&=&\int_{r=r_\infty} d^4 x\sqrt{-\gamma}\bigg(-6-|\Phi|^2+ |B_{\mu\nu}|^2+\frac{1}{2}(\log r^2)\bigg[
\frac{1}{4}\mathcal{F}^2+\frac{1}{4}F^2+|D_\mu\Phi|^2+\nn\\
&&
~~~~+\bigg(\frac{1}{3}+\frac{\lambda_1}{2}\bigg)|\Phi |^4
+\frac{1}{3\eta}\big(\mathcal{D}_{[\mu}B_{\alpha\beta]}\big)^*\big(\mathcal{D}^{[\mu}B^{\alpha\beta]}\big)-|B_{\mu\nu}|^4+ \lambda |\Phi |^2|B_{\mu\nu}|^2\bigg] \bigg)\nn
\eea
where $\gamma_{\mu\nu}$ is the metric induced by $g_{ab}$ on the boundary via $\gamma_{ab}=g_{ab}-n_a n_b$ where $n_a$ is outward pointing unit normal vector of the boundary. $K=\gamma^{ab}\nabla_a n_b$ is the trace of the extrinsic curvature with respect to the metric at the boundary. 
 \para
From the $zz$ component of the Einstein equation (\ref{eq:ein1}), we have $R_{zz}-\frac{1}{2}g_{zz}\mathcal{L}=0$. Thus the bulk on-shell action is a total derivative 
\be
S=\int d^4x dr \sqrt{-g}\mathcal{L}=-\int d^4 x \int _0^{r_\infty} dr\big[fu^{1/2}u'\big]'.
\ee
Taking into account the boundary terms and performing a Wick rotation, the free energy density can be obtained as 
\be\frac{\Omega}{V}
=\frac{7}{9}b^4-\frac{4bb_2}{3}+\frac{7M^4}{36}-3u_2+\frac{5+8\lambda}{9}b^2M^2-2 M\phi_2\,.
\ee
The stress tensor for the dual field theory can be calculated as 
\be
T_{\mu\nu}=2(K_{\mu\nu}-\gamma_{\mu\nu}K)+\frac{2}{\sqrt{-\gamma}}\frac{\delta S_\text{c.t.}}{\delta \gamma^{\mu\nu}}\,.
\ee
The total energy density is $\epsilon=\lim_{r\to\infty}\sqrt{-\gamma} \langle T^0_0\rangle=\frac{7}{9}b^4-\frac{4bb_2}{3}+\frac{7M^4}{36}-3u_2+\frac{5+8\lambda}{9}b^2M^2-2 M\phi_2$, thus $\frac{\Omega}{V}=\epsilon.$
 \para
The free energy can be found in the main text and we conclude that the system is smooth when it crosses the phase transition. 

\section{Probe limit}
\label{secab}

The simplest way to solve the system is to consider the probe limit $\kappa/L^{3/2} \ll 1$. However, the probe limit for the nodal line semimetal system is not well defined near the horizon and even with another set of scaling dimensions where the probe limit is well defined near the horizon it is not valid near the critical point.  For the axial vector field in the holographic Weyl semimetal, the probe limit is a good limit for the same scaling dimensions of operators dual to $A_{z}$ and $B_{xy}$ except near the critical point.
 \para
In the probe limit, we assume that the Newton coupling constant is very small so that the backreaction to the geometry could be ignored. We need to check if the probe limit is satisfied for a given solution of $B$ and $\phi$. The equations of motion for $B$ and $\phi$ in the $AdS_5$ background is
\bea
B''+\frac{B'}{r}-\frac{m_b^2B}{r^2}-\frac{\eta B \phi^2}{2 r^2}=0 \,,\\
\phi''+\frac{5\phi'}{r}-\frac{m_\phi^2\phi}{r^2}-\frac{\eta B^2\phi}{r^6}=0\,.
\eea
We have to make sure that the source term of $B$ and $\phi$ at both the boundary and the horizon are small enough not to cause too much backreaction. In the probe limit, the IR scaling dimention of $B$ and $\phi$ is the same as the UV scaling.  Thus it seems that there is only one scaling dimenion that we can use, which is zero for both fields, otherwise, the fields either backreact too much at the horizon or too much at the boundary. In this way, the two fields contribute at $\kappa^2$ order compared to the background and when $\kappa\to 0$ the probe limit is a physically well defined limit for finite solutions. 
 \para
Let us focus on $m_\phi=0$ and $m_B=2$. There are three types of near horizon solutions
The first solution is 
\bea
\phi&\simeq&\phi_0+\phi_1(\phi_0,b_1) r^{\sqrt{16+2\lambda \phi_0^2}-4}+\cdots,\\
B_{xy}&\simeq& b_1 r^{4+\lambda\phi_0^2/2}+\cdots,
\eea where $b_1$ is a tuning parameter and the second solution is
\bea
\phi&\simeq&\phi_1 r^{\sqrt{4+\lambda b_0^2}-2}+\cdots,\\
B_{xy}&\simeq& b_0r^2+b_1(b_0,\phi_1) r^{2(\sqrt{4+\lambda b_0^2}-1)}
\eea where $\phi_1$ is a tuning parameter.
The critical solution is $B_{xy}=1/\sqrt{2}r^2\phi$ and the critical point is $M/b=\sqrt{2}$. This is a special property of the probe limit at the critical point, which is also true for the holographic Weyl semimetal model at the probe limit. Flowing this geometry to the boundary we find that at the boundary, though the scaling dimension of $B$ and $\phi$ guarantees that free $B$ or $\phi$ does not have backreaction in the probe limit, the interaction at the UV makes both $B$ and $\phi$ more divergent than their scaling dimensions, which means that it cannot flow to asymptotic $AdS_5$ solutions. 
 \para
To solve this problem and make the probe limit well defined, we introduce another scalar field $\lambda$ which mediates the interaction between $B$ and $\phi$ and choose a scaling dimension for $\lambda$ so that the interaction term is not important in the UV so that the probe limit is still valid. We change the interaction term of $\phi$ and $B$ to $\lambda^2 B^2\phi^2$ and introduce the kinetic term for $\lambda$ in the action with a relative minus sign. The equations of motion now become
\bea
B''+\frac{B'}{r}-\frac{m_b^2B}{r^2}-\frac{\eta\lambda^2 B \phi^2}{2 r^2}=0\, ,\\
\phi''+\frac{5\phi'}{r}-\frac{m_\phi^2\phi}{r^2}-\frac{\eta\lambda^2 B^2\phi}{r^6}=0\,,\\
\lambda''+\frac{5\lambda'}{r}-\frac{m_{\lambda}\lambda}{r^2}-\frac{V_0 \lambda^3}{r^2}+\frac{\eta\lambda B^2\phi^2}{r^6}=0\,,
\eea where $B$ denotes $B_{xy}$.
We choose $m_{\lambda}^2=-3$ so that the interaction at the UV is not important. For the interaction to be not important in the IR we introduce a $\lambda^4$ potential term for the $\lambda$ scalar.
 \para
The near horizon boundary conditions now have a new field $\lambda=\sqrt{3/V_0}+\cdots$ where the subleading terms in $\lambda$ depends on the phase of the solution. For the other two fields, the near horizon boundary condition does not change except to substitute $\eta$ with $3\eta/v_0 $. Flowing these solutions to the boundary and we could also get a valid probe limit system with three different types of solutions. This probe limit is well defined for most of the parameter regime. However, near the critical point the probe limit becomes subtle because the solutions become larger and larger near the boundary when getting more and more close to the critical point.  Thus considering backreactions is a better and more physical choice, and here we do not elaborate more on this probe limit.

 \end{document}